\documentclass[a4paper,12pt]{article}
\usepackage[utf8]{inputenc}
\usepackage[english]{babel}
\usepackage{amsmath}
\usepackage{authblk}
\usepackage{graphicx}
\usepackage[singlespacing]{setspace}
\usepackage[headheight=1in,margin=1in]{geometry}
\usepackage{fancyhdr}
\usepackage{lipsum}
\usepackage{hyperref}
\usepackage{xcolor}
\usepackage{amsfonts}
\usepackage{amsmath}
\usepackage{latexsym}
\usepackage{caption}
\usepackage[labelfont=bf]{caption}
\usepackage{makecell}
\usepackage{url}
\usepackage{subcaption}

\usepackage{caption}
\usepackage{natbib}
\setcitestyle{authoryear,open={(},close={)}} 

\date{}

\begin{document}
\title{Occupational Convergence or Divergence?\\
Mapping Labor Market Structural Shifts \\Driven by AI Penetration}
\author{
Rafiazka Hilman,\textsuperscript{1,2}, Júlia Koltai\textsuperscript{1,3}
}

\maketitle
\begin{center}
{
1. MTA–TK Lendület “Momentum” Digital Social Science Research Group for Social Stratification, ELTE Centre for Social Sciences, Budapest, Hungary\\
2. Informatics Institute, University of Amsterdam, Amsterdam, The Netherlands\\
3. Institute of Empirical Studies, ELTE Faculty of Social Sciences, Budapest, Hungary
}\\
\end{center}

\vspace{10mm}

\begin{abstract}
\fontsize{9pt}{11pt}\selectfont 
Artificial intelligence (AI) is rapidly becoming a defining feature of contemporary labor markets, yet it remains unclear whether its diffusion is producing a common set of competencies across occupations or deepening occupational divisions. In this study, we investigate how AI related skill demand is reshaping labor market structure using large-scale online vacancy data from ten countries spanning the Global North and Global South. Combining natural language processing and a large language model with multilevel bipartite network analysis, we map the relationships between occupations, required skills, and career stages in the emerging AI economy. We find that AI demand is overwhelmingly concentrated within a narrow technical core, with approximately three quarters to four fifths of AI related vacancies located in STEM occupations across all countries. At the same time, AI intensive jobs share a remarkable focus on Python, SQL, machine learning, and data analysis, generating convergence among highly exposed occupations. Interestingly, this convergence does not extend across the wider labor market. Instead, AI competencies remain largely confined to technical domains and are most strongly demanded at the point of labor market entry, where they increasingly function as prerequisites for occupational access. These findings reveal a bifurcated pattern of technological change namely convergence within an already advantaged AI exposed core and divergence between that core and the rest of the occupational structure. Rather than broadly democratizing opportunities, AI appears to reorganize labor demand along existing lines of occupational stratification, potentially raising barriers to entry and concentrating the benefits of AI adoption among workers and occupations with prior technological advantages. Our results suggest that the inequalities associated with AI are embedded not only in employment outcomes but also in the evolving architecture of skill demand itself.
\end{abstract}

\clearpage

\section{Introduction}
\label{Introduction}
The increasing penetration of artificial intelligence (AI) is transforming the nature of work across a wide range of occupational domains. As organizations integrate AI into production processes, decision making systems, and service delivery, workers are increasingly expected to interact with AI enabled tools and technologies in their daily professional activities. Consequently, labor markets are experiencing a gradual reconfiguration of skill requirements with AI related competencies becoming more visible in job advertisements and recruitment practices. While the diffusion of AI has generated considerable discussion regarding its implications for productivity and employment, less attention has been devoted to understanding how AI related skill demands are reshaping the broader structure of occupations. In particular, an important question remains whether the growing demand for AI competencies is fostering equally similarly across occupations or the required new skills are unequally distributed across fields.

Existing research suggests that the adoption of AI is unevenly distributed across sectors, firms, and workers, resulting in heterogeneous patterns of adaptation and skill acquisition \citep{choi2024artificial, zhang2025automation}. At the same time, the increasing appearance of AI related requirements in job advertisements indicates that employers are revising fitting qualification profiles to accommodate new technological capabilities \citep{chen2025artificial}. However, the implications of these changes for occupational structure remain insufficiently understood. Although prior studies have documented the emergence and economic value of AI skills \citep{stevenson2018economics}, they have largely focused on questions of adoption \citep{liu2024embracing}, productivity \citep{georgieff2022artificial}, and wage outcomes \citep{felten2019occupational} rather than on how AI related competencies are distributed in the labor market across occupations. As a result, a critical gap persists concerning whether AI skill demand contributes to occupational convergence of the different labor market segments by creating a shared set of competencies across job categories, or occupational divergence of the labor market segments by increasing the specialization and distinctiveness of occupational skill profiles.

Addressing this gap is important because the answer has direct implications for workforce mobility, occupational structures, and the future organization of labor markets. If AI related competencies diffuse broadly across occupations, they may function as a common skill layer that facilitates transitions between jobs and increase occupational and social mobility. Conversely, if required AI skills are concentrated within particular occupational clusters and combined with highly specialized knowledge, they may reinforce existing divisions and generate new forms of labor market stratification which make the occupational and social mobility harder in the future. Moreover, the growing economic value attached to AI competencies underscores the significance of understanding their distribution. Job postings that require AI related skills are associated with substantial wage premiums \citep{felten2019occupational}, both within firms and within comparable job titles, while firms with stronger AI skill requirements tend to offer higher compensation overall \citep{alekseeva2021demand}.

Building on these observations, this study examines the relationship between AI skill demand and occupational structure on an international dataset using large scale job listing data, providing a cross national perspective on how AI adoption is reshaping labor market demand across countries in both the Global North and the Global South. To investigate these transformations, we pursue two primary objectives. First, we examine the relationship between required work experience and the demand for AI related skills across occupations in order to identify which career stages are most likely to be affected by AI adoption. Second, we explore the structural interdependencies between job categories and AI competencies through a bipartite network framework that captures the organization of emerging labor market requirements.

Our ultimate aim is to introduce a comprehensive perspective on the changing role of AI within labor market and the ways in which technological adoption reconfigures occupational demands. By developing network-based approaches, this study highlights regional, occupational, and career stage variations in AI skill demand and provides new insights into the uneven diffusion of AI related competencies in contemporary labor markets.\\

\subsection{Structural Transformation of the Labor Market}
\label{Structural Transformation of the Labor Market}
The diffusion of AI is increasingly recognized as a transformative force in contemporary labor markets, extending beyond firm level productivity gains to influence the broader structure of employment. Unlike previous waves of technological change that primarily automated routine physical tasks, AI possesses the capacity to perform a growing range of cognitive and analytical functions, altering both the composition of occupational demand and the nature of work across industries and sectors \citep{frank2019toward}. As AI technologies become integrated into production systems, organizational processes, and service delivery, labor markets are experiencing shifts in the relative value of different skills, occupations, and forms of human capital \citep{ernst2024artificial}. These developments have stimulated growing scholarly interest in understanding how AI contributes to structural changes in employment patterns and occupational structures.

The relationship between AI and labor market transformation is frequently examined through the lens of task transformation. Rather than affecting entire occupations uniformly, AI technologies alter the demand for specific tasks that constitute occupational activities. As organizations increasingly adopt AI-driven systems, some tasks become automated while others are enhanced through technological assistance. This distinction between automation and augmentation has emerged as a central framework for understanding how AI reshapes employment opportunities, skill requirements, and workforce inequalities. Consequently, the labor market effects of AI are often uneven, varying according to the nature of occupational tasks and the competencies required to perform them.

The automation of routine and standardized tasks has generated concerns regarding the displacement of workers in occupations characterized by repetitive cognitive and administrative activities. AI systems are increasingly capable of performing tasks involving repetitive information processing, administrative procedures, and predictable decision making. Despite automating these tasks, AI technologies, as a form of innovation, may exert an overall positive effect on employment \citep{damioli2023ai}. At the same time, AI does not solely function as a substitute for labor. In many occupations, AI acts as a complementary technology that enhances human productivity and expands the scope of professional activities. Through augmentation, workers can leverage AI systems to improve problem solving, data analysis, communication, and decision-making processes. Occupations requiring advanced analytical, technical, or creative capabilities are particularly likely to benefit from these complementarities \citep{marguerit2025augmenting}. As firms increasingly adopt intelligent technologies, demand shifts toward workers possessing skills that complement AI systems, while workers performing highly automatable tasks face greater risks of displacement and declining labor market prospects.

These uneven adjustments in task demand contribute to broader patterns of skill polarization and occupational restructuring. One prominent stream of the literature argues that AI disproportionately affects occupations located in the middle of the skill distribution, where routine cognitive activities are more susceptible to automation. Consequently, demand for many medium skill occupations declines, while employment opportunities expand in occupations requiring advanced analytical, technical, and managerial capabilities. This divergence creates a more polarized employment structure characterized by unequal access to opportunities, earnings, and career advancement, reinforcing existing labor market inequalities \citep{hampole2025artificial}. These dynamics may be particularly consequential in developing economies, where labor markets often contain larger concentrations of occupations vulnerable to automation and where limited access to advanced education, digital infrastructure, and workforce retraining programs constrains workers' capacity to adapt to evolving technological requirements \citep{ganuthula2025skill}.

Beyond reshaping existing occupations, AI also creates new occupations, tasks, and forms of work that did not previously exist, generating opportunities for labor market renewal and adaptation \citep{sarikaya2025artificial}. The widespread deployment of AI has given rise to roles such as AI ethics officers and responsible AI auditors, who oversee governance, compliance, and risk management, while occupations including AI trainers, prompt engineers, and AI Ops specialists have emerged to develop, deploy, and maintain AI systems \citep{SrinivasanChenZakerinia2026HBS}. These examples illustrate how technological advances not only transform existing jobs but also generate new forms of specialized work within the evolving AI ecosystem.

At the macroeconomic level, these transformations contribute to the structural reallocation of labor across occupations and industries. AI enables firms to reorganize production processes, improve operational efficiency, and develop new products and services, thereby accelerating industrial upgrading and productivity growth \citep{acemoglu2018artificial, stevenson2018economics}. As industries adopt increasingly sophisticated technologies, demand shifts toward workers capable of complementing AI systems, while occupations characterized by automatable tasks become relatively less important. These competing dynamics highlight the dual nature of AI-driven technological change with processes of job destruction and job creation occur simultaneously, often affecting different groups of workers in uneven ways. Consequently, these collective forces reshape occupational structures, transform skill demand, and contribute to the broader structural transformation of contemporary labor markets \citep{yan2024evaluating}.\\

\subsection{Geographical and Social Inequalities in the Labor Market}
\label{Geographical and Social Inequalities in the Labor Market}
The impacts of artificial intelligence (AI) on labor markets are not distributed uniformly across geographical regions. The adoption of AI technologies varies considerably between regions due to differences in their industrial composition, technological infrastructure, educational attainment, innovation capacity, and labor force characteristics. Consequently, the effects of AI on employment and occupational change are often shaped by local economic conditions and regional development trajectories. Accordingly, understanding these spatial dimensions is important because technological transformation occurs within geographically embedded labor markets, across which opportunities and vulnerabilities are distributed unevenly.

Existing research indicates that regions with higher levels of AI adoption frequently experience significant shifts in employment structures \citep{carbonero2023impact}. While AI can contribute to productivity growth and economic modernization, its diffusion may also reduce labor demand in occupations characterized by tasks that are susceptible to automation. Evidence suggests that regions with greater AI exposure tend to exhibit declines in employment-to-population ratios, particularly among workers employed in middle skill and non-STEM occupations \citep{huang2024labor}. These findings imply that the benefits of technological progress may not be shared equally across occupational groups and regions. Instead, AI adoption may intensify existing disparities between locations that possess strong capacities for technological adaptation and those whose labor markets remain concentrated in occupations vulnerable to displacement.

Beyond regional variation, AI driven labor market transformation also exhibits important demographic inequalities. Exposure to AI technologies differs across social groups due to their interdependence with educational attainment, occupational specialization, sectoral employment, and access to digital resources. As a result, some groups may be better positioned to benefit from AI related opportunities, while others face heightened risks of labor market disruption \citep{marguerit2025augmenting}. Recent evidence highlights the existence of substantial spatial and gender differences in AI exposure, suggesting that technological change may reinforce preexisting inequalities in employment outcomes and career opportunities \citep{cranney2026global}. These patterns demonstrate that the consequences of AI adoption extend beyond occupational restructuring and encompass broader questions of social and economic inequality. From a network perspective, such spatial and demographic disparities reflect the uneven diffusion of AI related skills and competencies across labor markets \citep{acemoglu2018artificial}. Regions and occupational groups are connected through flows of knowledge, workers, and technological capabilities, creating complex networks through which AI related demands spread and evolve. 
\\

\section{Data and Methods}
For the exploration on how AI is reshaping labor market structure, we developed an integrated pipeline comprising data collection, computation, and visualization (Figure \ref{fig1}). The data collection stage employs a keyword based web scraping strategy, detailed in Supplementary Materials B.1, in which job advertisements are retrieved from major online recruitment platforms using a predefined dictionary of AI related terms. The dictionary was generated by prompting GPT-OSS through the Ollama framework to produce a comprehensive vocabulary of AI related skills, tools, methodologies, and technologies commonly appearing in job advertisements (e.g.: machine learning, deep learning, natural language processing, automation, and predictive analytics). This keyword based retrieval strategy ensures that the collected dataset focuses on vacancies requiring AI-related competencies while maximizing the coverage of relevant advertisements across occupations, industries, and seniority levels. The scraped job descriptions are subsequently processed using two-stage information extraction (2SIE) approach, a hybrid Natural Language Processing (NLP) framework combining rule-based mapping (tokenization, lemmatization, and keyword matching) with prompt based information extraction to improve the accuracy and robustness of structured information extraction. The rule based stage standardizes the textual content while the prompt-based stage extracts structured information from job advertisements, resulting in a consolidated classification with optimized metadata for subsequent data cleaning. Stored in a web hosting container, this ready-to-use database is maintained as a static comma-separated values (CSV) file to ensure interoperability during computation and automated visualization streaming from the backend to the webpage for dashboard deployment. The detail description is available in Supplementary Materials B.2.

\begin{figure}[!ht]
  \centering
\includegraphics[width=0.9\textwidth]{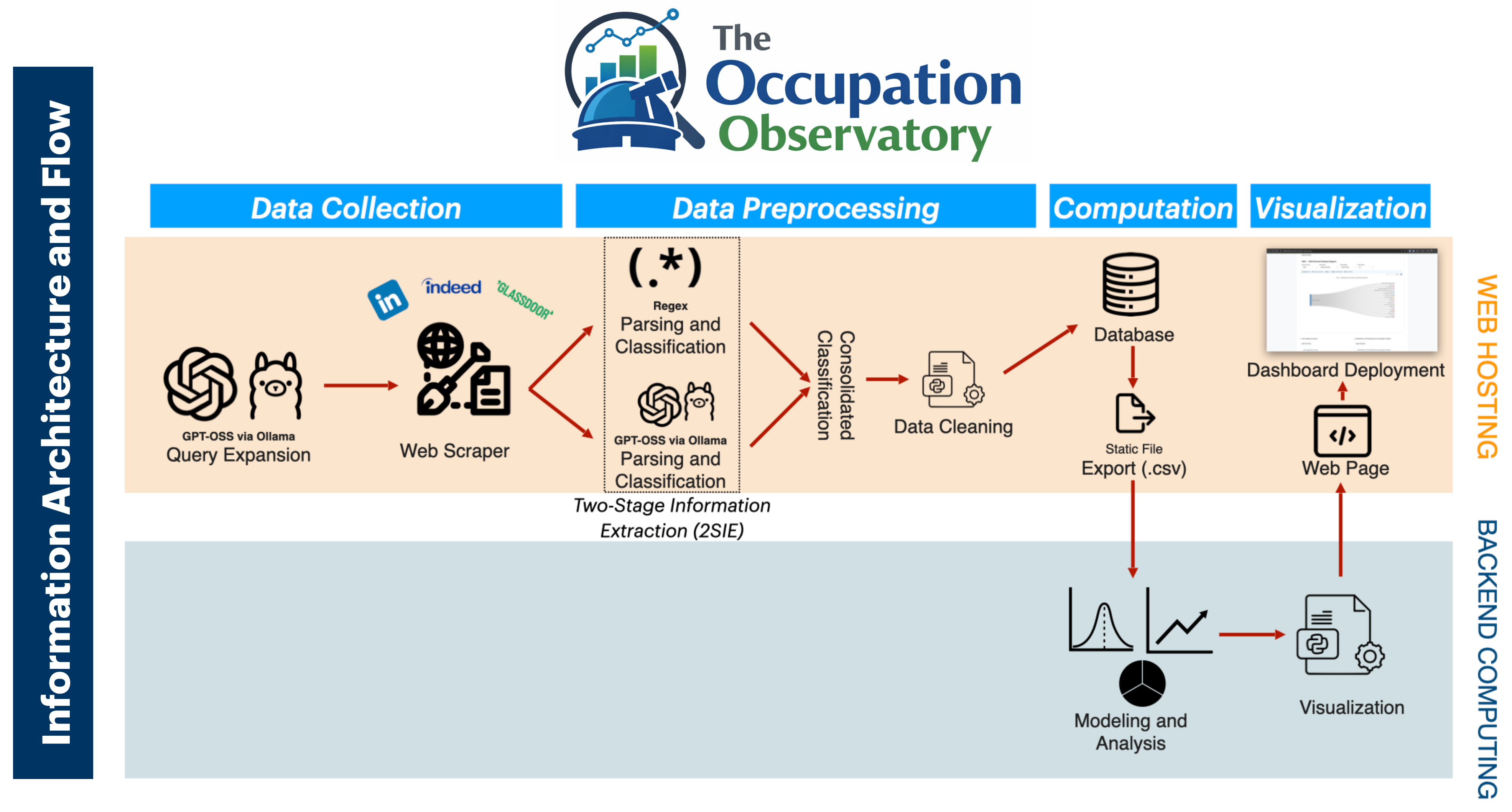}
  \caption{Pipeline design and two-stage information extraction (2SIE) approach. Job postings scraped from LinkedIn, Indeed, and Glassdoor are processed through four stages: i) data collection; ii) data preprocessing; iii) computation, and iv) visualization. The 2SIE combines NLP and regex parsing with zero-shot LLM classification (GPT-OSS via Ollama) to extract standardized skill, seniority, and SOC information for downstream network analysis and the interactive Occupation Observatory dashboard.}
\label{fig1}
\end{figure}

Examination of the distribution of AI related keywords is then performed across SOC categories, highlighting which occupational groups demand AI intensive roles more frequently and how these roles correlate with years of experience. The whole pipeline ends with an interactive dashboard namely \textit{'The Occupation Observatory'}, which enables users to explore labor market insights through thematic visualizations. It displays AI skill demand by occupation category and country, geographic distributions of job listings, and word clouds of key competencies. A bipartite network on the dashboard maps relationships between occupations, skills, and seniority levels, while cluster analysis highlights how experience intersects with AI skill requirements.

\subsection{Data}
\label{Data}
At first, web extraction is conducted to scrape structured data, including job titles, job descriptions, listed qualifications, posting dates, and other metadata, from LinkedIn, Indeed, and Glassdoor between December 2025 and May 2026. The keyword-based filtering strategy targets advertisements related to Artificial Intelligence, Data Science, Data Analytics, and Business Intelligence, using a predefined dictionary of AI related terms expanded through prompt engineering with GPT-OSS deployed via the Ollama environment as detailed in Supplementary Materials B.1. The geographical coverage of this study includes both the Global North and the Global South, encompassing ten countries: Brazil, China, Germany, India, Indonesia, the Netherlands, Singapore, South Africa, the United Kingdom, and the United States.
\\

\subsection{Data Parsing and Classification via Prompt Engineering}
\label{Skill Classification via Prompt Engineering}
To classify skills from job advertisements, we employed a zero-shot prompt engineering strategy. This prompting technique enables large language models to perform task-specific classification and information extraction without requiring labeled examples, relying instead on natural language instructions that specify the desired task and output format \citep{brown2020language, kojima2022large}. The prompt instructed the language model as seen in in Supplementary Materials B.2 to act as an expert in job analysis and to extract a standardized set of information from each job description, including expected technical skills, expected soft skills, the last submission date when available, and the most appropriate SOC code. In addition, the prompt required the model to identify whether any extracted technical skills were related to artificial intelligence, data science, or data analytics, and to generate a concise English summary of the job description.

The zero-shot prompt was designed to minimize ambiguity while preserving consistency across job postings. From the technical side, the model was constrained to return a structured JSON object containing six fields: 'expected\_skills', 'has\_ai\_ds\_skill', 'expected\_softskills', 'last\_submission\_date', 'soc\_label', and 'description\_en'. Technical and soft skills were translated into English, when necessary, prior to classification and incorporated into the prompt described in Supplementary Materials B.2, dates were standardized to 'YYYY-MM-DD' format, and missing information were coded as null. Occupational labels were assigned according to the SOC system, a hierarchical framework used to classify workers and jobs into standardized occupational categories based on the work performed \citep{bls2018soc}. Because the SOC taxonomy was explicitly embedded in the prompt, this approach enabled efficient large scale classification of job postings without requiring task-specific examples while ensuring that the outputs remained machine readable and directly suitable for downstream analysis.

\subsection{Skill Rank Space}
\label{Skill Rank Space}
To characterize the concentration of skill demand, we construct a
\emph{Skill Rank Space} that orders competencies by their prevalence and examines how demand is distributed across occupational groups and countries. The
skill rank space summarizes the marginal distribution of demand for individual competencies, making assessment on whether AI related skill demand is concentrated in a small core of dominant competencies or dispersed across a broad set possible.

For a given occupation cluster group $g$, computed separately for each country, let $n_g(s)$ denote the number of job postings in $g$ that require skill $s$, and let $N_g$ be the total number of postings in $g$. The demand share of skill $s$ is

\begin{equation}
f_g(s) = \frac{n_g(s)}{N_g},
\label{eq:demand-share}
\end{equation}

which gives the proportion of postings within the cluster that request skill $s$.

Skills are then sorted in descending order of demand share and assigned an integer
rank $\rho_g(s) \in \{1, 2, \dots, \lvert S_g \rvert\}$ such that

\begin{equation}
f_g\!\left(s_{(1)}\right) \ge f_g\!\left(s_{(2)}\right) \ge \cdots \ge
f_g\!\left(s_{(\lvert S_g \rvert)}\right),
\label{eq:rank-order}
\end{equation}

so that $\rho = 1$ identifies the most frequently demanded competency. The skill rank space is the set of coordinate pairs $\{(\rho_g(s),\, f_g(s))\}$, which
we visualize as a rank frequency curve with the leading skills annotated through
superimposed word clouds in which type size is proportional to $f_g(s)$. Accordingly, the shape of the rank frequency curve characterizes the concentration of skill demand. A steeply declining, heavy tailed curve implies that a few competencies account for the majority of demand, whereas a flatter curve implies more even dispersion.

\subsection{Bipartite Network Projection}
\label{Bipartite Network Projection}

\subsubsection{Bipartite Network of Occupations and Skills}
To examine the structural relationships between occupations and skills, we model the labor market as a bipartite network $G=(U,V,E)$, where $U$ represents occupation related nodes and $V$ represents skill nodes. The occupation layer comprises three hierarchical categories namely occupation groups, occupations, and seniority levels, while an edge $e_{ij} \in E$ connects an occupation related node $u_i$ to a skill node $s_j$ whenever the skill is required in a job posting. Bipartite representations are particularly suitable for labor market analysis because they preserve the original relationships between occupations and competencies without imposing assumptions about direct connections among occupations or among skills \citep{newman2001scientific, zhou2007bipartite}.

Formally, the bipartite graph is defined as

\begin{equation}
G=(U,V,E),
\end{equation}

where

\begin{equation}
U = U_m \cup U_o \cup U_s,
\end{equation}

and $U_m$, $U_o$, and $U_s$ represent occupation groups, occupations, and seniority levels, respectively. The set of skill nodes is denoted by
\begin{equation}
V=S,
\end{equation}

and the set of edges satisfies

\begin{equation}
E \subseteq U \times V.
\end{equation}

An edge exists whenever occupation related node $u_i$ requires skill $s_j$,

\begin{equation}
e_{ij}=
\begin{cases}
1, & \text{if } u_i \text{ requires } s_j,\\
0, & \text{otherwise}.
\end{cases}
\end{equation}

The network can be represented by a bipartite incidence matrix $B=[b_{ij}]$, where rows correspond to occupation related nodes and columns correspond to skills. To account for differences in skill prevalence, we construct a weighted bipartite network in which edge weights represent the relative frequency with which a skill is required within an occupation related category. Specifically, the weight $w_{ij}$ is calculated as

\begin{equation}
w_{ij} =
\frac{n(u_i,s_j)}
{N(u_i)},
\end{equation}

where $n(u_i,s_j)$ denotes the number of job advertisements associated with occupation related category $u_i$ that require skill $s_j$, and $N(u_i)$ denotes the total number of job advertisements assigned to occupation related category $u_i$. Consequently, $w_{ij}$ represents the proportion of advertisements within occupation related category $u_i$ that require skill $s_j$. This weighting approach allows direct comparison of skill prevalence across occupation related categories with different numbers of advertisements while preserving the relative importance of each skill.

\subsubsection{Projected Network of Occupations}
We then generate the one-mode projection of the bipartite network. In the occupation projection, two occupation nodes are connected if they share one or more AI related skills, producing a network that captures occupational similarity based on common AI skill requirements. Edge weights in the occupation projection are computed as the number of shared AI related skills between each pair of occupations using the unweighted version of the original bipartite network. The resulting edge weight between occupations $u_i$ and $u_j$ is given by

\begin{equation}
\sum_{k=1}^{|V|}
b_{ik}b_{jk},
\end{equation}

which measures the number of skills shared by the two occupation related nodes. Such projections are widely used in network analysis to uncover latent structures and interdependencies embedded within two mode systems \citep{newman2001scientific, zhou2007bipartite}.
To facilitate comparison between occupations with different numbers of required skills, we additionally calculate the Jaccard similarity coefficient,

\begin{equation}
\frac{|S_i \cap S_j|}
{|S_i \cup S_j|},
\end{equation}

where $S_i$ and $S_j$ denote the sets of skills associated with occupations $u_i$ and $u_j$, respectively. Higher values indicate greater overlap in competency requirements and therefore stronger occupational similarity.

To evaluate the prominence of occupations within the projected networks, we calculate node strength as

\begin{equation}
\sum_{j=1}^{N}
w_{ij},
\end{equation}

where $N$ denotes the number of neighboring nodes and $w_{ij}$ represents the corresponding edge weight. Node strength captures the cumulative intensity of occupation relationships and identifies the most influential occupations within the network \citep{newman2001scientific}.

Finally, the resulting projected networks provide the basis for evaluating whether AI related competencies contribute to occupational convergence or divergence. On one hand, occupation projections reveal the extent to which different occupations share common AI skill requirements. To assess convergence quantitatively, we compute the average occupational similarity across all occupation pairs,

\begin{equation}
\frac{1}{M}
\sum_{i<j}
J_{ij},
\end{equation}

where $M$ denotes the total number of occupation pairs in the projected network. Higher values of $C$ indicate greater overlap in occupational skill requirements and thus stronger occupational convergence, whereas lower values suggest increasing specialization and occupational divergence. To improve robustness, network interpretation follows recommendations from statistically validated projection methods that minimize spurious links generated by highly connected nodes \citep{saracco2017inferring}.

\subsection{Occupational Clustering}
\label{Occupational Clustering}
To identify groups of occupations that share similar skill requirements, we perform community detection on the projected occupation networks generated from the bipartite occupation–skill structure. Community detection enables the identification of densely connected groups of nodes whose members exhibit stronger internal similarity than external similarity. In the context of labor markets, these communities represent clusters of occupations that require comparable skill portfolios and therefore occupy similar positions within the occupational structure. Because the occupation network is analyzed at multiple levels of aggregation, including occupation groups, occupations, and seniority categories, the clustering procedure is applied separately to each projected network, enabling the examination of occupational convergence and divergence across different scales of labor market organization \citep{fortunato2010community, valejo2017onemode}.
As the primary clustering approach, we employ the Louvain algorithm, which partitions a network by maximizing modularity, a measure of the extent to which observed connections within communities exceed those expected under a random network model \citep{blondel2008fast}. Modularity is defined as

\begin{equation}
Q
=
\frac{1}{2m}
\sum_{i,j}
\left(
A_{ij}
-
\frac{k_i k_j}{2m}
\right)
\delta(c_i,c_j)
\end{equation}

where $A_{ij}$ denotes the weighted adjacency matrix of the projected network, $k_i$ and $k_j$ represent the weighted degrees of nodes $i$ and $j$, $m$ is the total edge weight in the network,

\begin{equation}
\frac{1}{2}
\sum_{i,j}
A_{ij},
\end{equation}

$c_i$ denotes the community assignment of node i, and

\begin{equation}
\delta(c_i,c_j)=
\begin{cases}
1, & \text{if } c_i=c_j,\\
0, & \text{otherwise}.
\end{cases}
\end{equation}

Higher values of $Q$ indicate stronger community structure and greater separation between occupational clusters \citep{newman2006modularity}.

To improve cluster quality and address limitations associated with disconnected communities, we additionally implement the Leiden algorithm. Leiden extends Louvain by introducing a refinement phase that guarantees well connected communities while preserving the objective of modularity optimization \citep{traag2019louvain}. The generalized modularity function is expressed as

\begin{equation}
Q=
\frac{1}{2m}
\sum_{i,j}
\left[
A_{ij}
-
\gamma \frac{k_i k_j}{2m}
\right]
\delta(c_i,c_j)
\end{equation}

where 
$\gamma$ is a resolution parameter controlling the granularity of the detected communities. Larger values of $\gamma$ yield smaller and more specialized clusters, whereas lower values generate larger and more aggregated communities. This flexibility is particularly useful for examining occupational structure at multiple levels of aggregation \citep{traag2019louvain}.

As a robustness check, we compare the community structure obtained from the Louvain algorithm with that identified by the Girvan-Newman algorithm, which detects communities through the iterative removal of edges with high betweenness centrality \citep{girvan2002community}. The purpose of this comparison is not to evaluate the relative performance of the two algorithms or to determine which partition is preferable, but rather to assess whether the principal community structure remains consistent across alternative community detection procedures. A high degree of agreement between the resulting partitions provides additional confidence that the identified occupational communities are not artifacts of a particular algorithmic implementation. Edge betweenness centrality is defined as

\begin{equation}
\sum_{s \neq t}
\frac{\sigma_{st}(e)}
{\sigma_{st}},
\end{equation}

where $\sigma_{st}$ denotes the total number of shortest paths between nodes $s$ and $t$, and $\sigma_{st}(e)$ represents the number of those paths that pass through edge $e$. Edges with high betweenness values frequently serve as bridges between communities. Their removal progressively reveals the underlying cluster structure of the network \citep{girvan2002community}.
Following community detection, we evaluate the characteristics of each occupational cluster using network-level indicators. The average internal similarity of a community $C_k$ is calculated as

\begin{equation}
\frac{2}
{|C_k|(|C_k|-1)}
\sum_{i<j}
J_{ij},
\end{equation}

where $J_{ij}$ denotes the Jaccard similarity between occupations $i$ and $j$ and $|C_k|$ is the number of occupations within community $k$. Higher values indicate stronger similarity in skill requirements among occupations belonging to the same cluster. 

\section{Results}
The empirical analysis proceeds in four stages. We first characterize the AI penetration in the labor market with broad occupation groups separately in each country, then decompose this demand into its constituent competencies at the sectoral level. We subsequently examine the structure of the occupational projected network at three levels of aggregation respectively seniority (macroscopic), SOC occupation group (mesoscopic), and individual occupation (microscopic). We interpret the observed patterns with respect to the study's central question of whether AI related skill demand fosters occupational convergence, by establishing a common competency layer across job categories, or occupational divergence, by reinforcing specialization and distinctiveness.

\begin{figure}[!ht]
  \centering
  \includegraphics[width=1\textwidth]{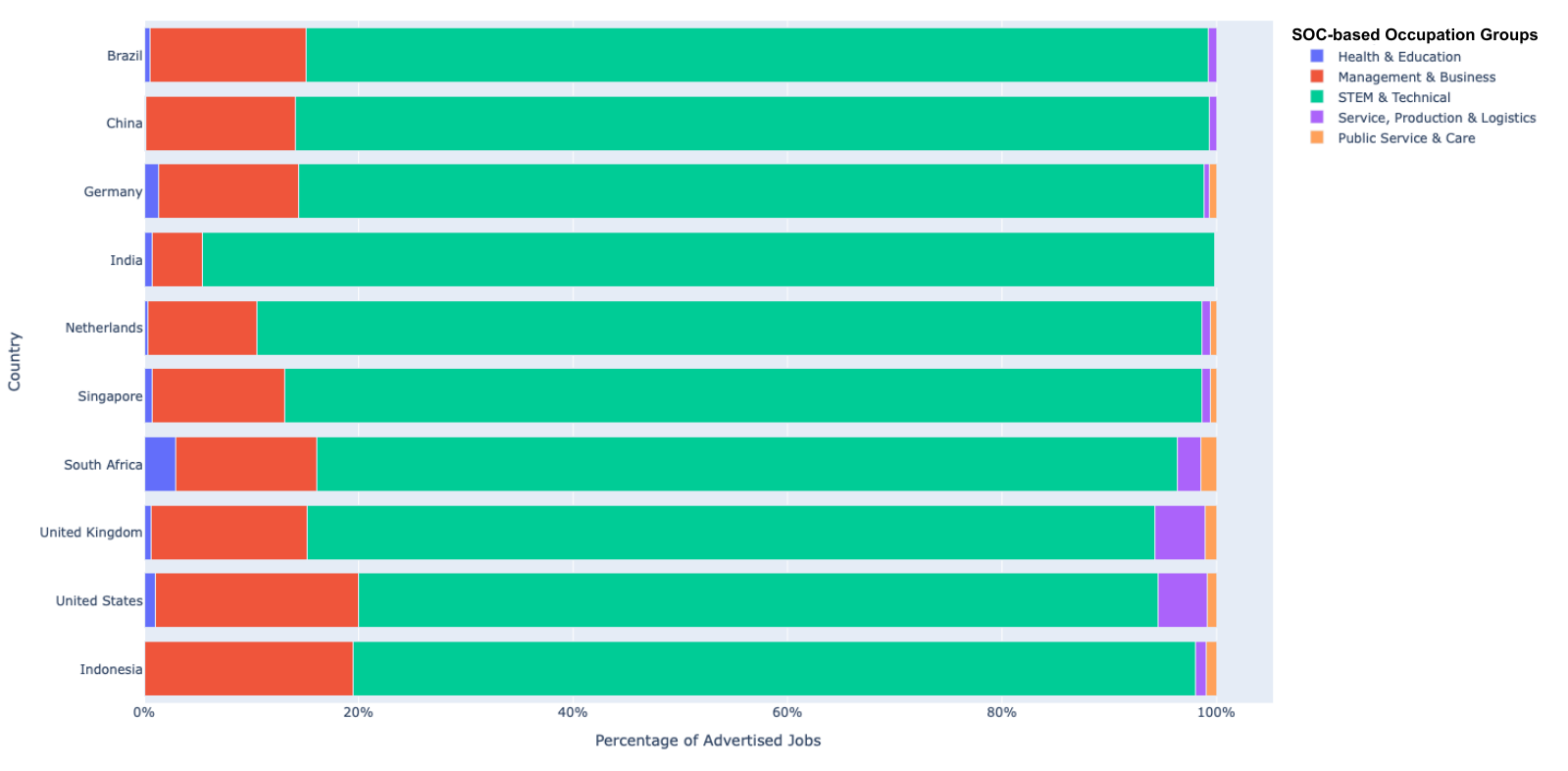}
  \caption{Distribution of SOC occupation groups among all advertisement including AI related skills by country. Stacked horizontal bars give the share of the five SOC-based occupation groups within all AI related job postings  for ten countries across the Global North and South where STEM \& Technical occupations domination is clearly present.}
\label{fig2}
\end{figure}

Figure \ref{fig2} shows the distribution of the five SOC occupation groups within all AI related job postings for the ten study countries. Across all national labor markets, demand is  concentrated in STEM \& Technical occupations, which account for roughly 75–85\% of AI related postings and reach their highest share in India. Management \& Business occupations form the second-largest group at approximately 10–20\% of postings, with a comparatively larger share in the United States and Indonesia than in India. The remaining clusters namely Health \& Education, Service, Production \& Logistics, and Public Service \& Care, each account for less than 5\% of AI related postings. 

Two features of this distribution are notable. First, the concentration of technical occupations among all AI related job post is strikingly consistent across both Global North and Global South economies, with only modest cross-national variation, indicating that the technical bias of AI demand is a global rather than a country specific phenomenon. Second, the limited penetration into non-technical sectors suggests that, at the level of broad SOC occupational groups, AI related skill demand has not yet diffused into a general-purpose requirement but remains bounded within a technical core, an early indication of between sector divergence rather than broad based convergence.

\begin{figure}[htbp]
    \centering
    \begin{subfigure}[b]{0.6\textwidth}
        \centering
        \includegraphics[width=\linewidth]{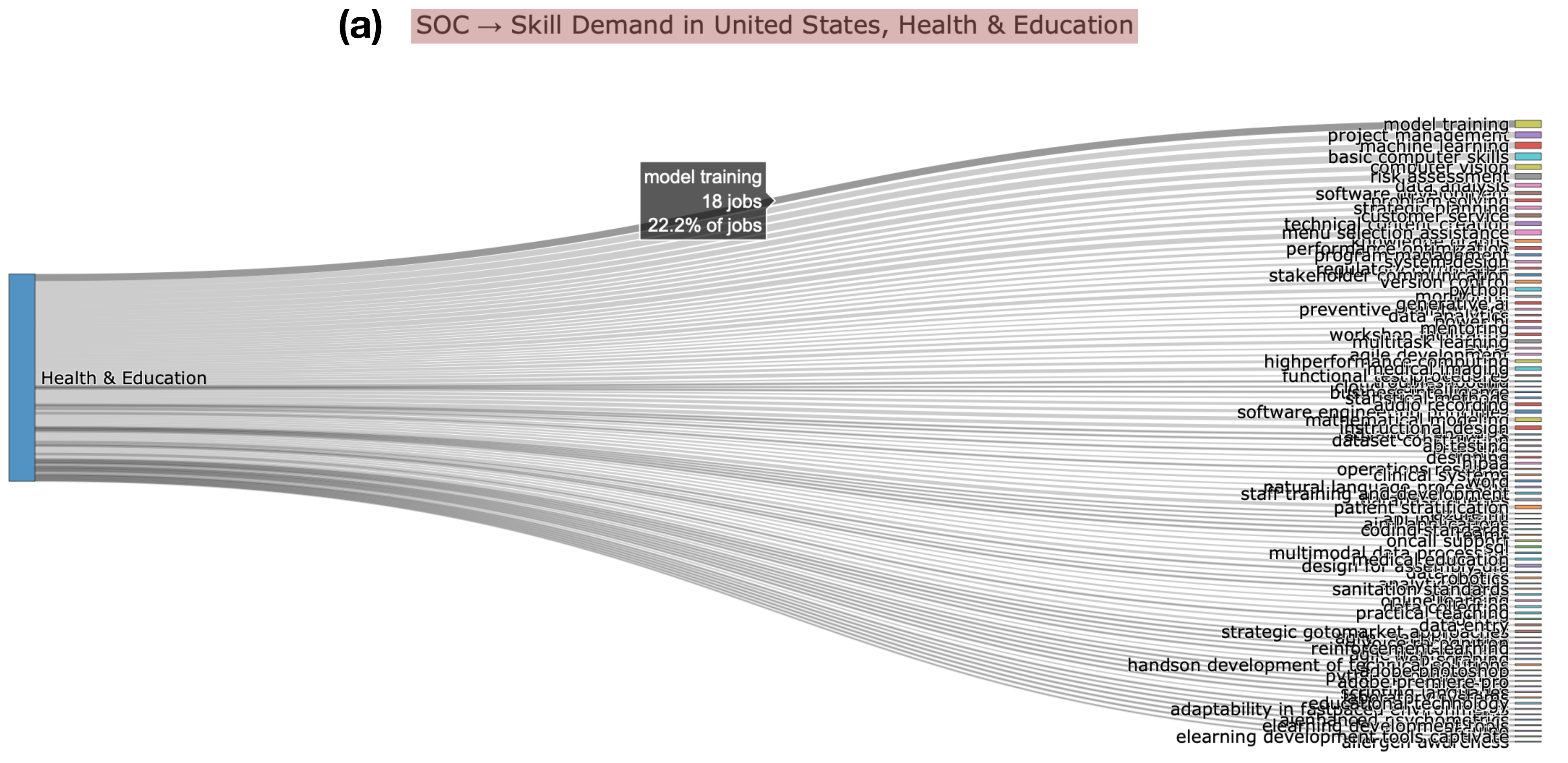}
    \end{subfigure}
    \hfill
    \begin{subfigure}[b]{0.6\textwidth}
        \centering
        \includegraphics[width=\linewidth]{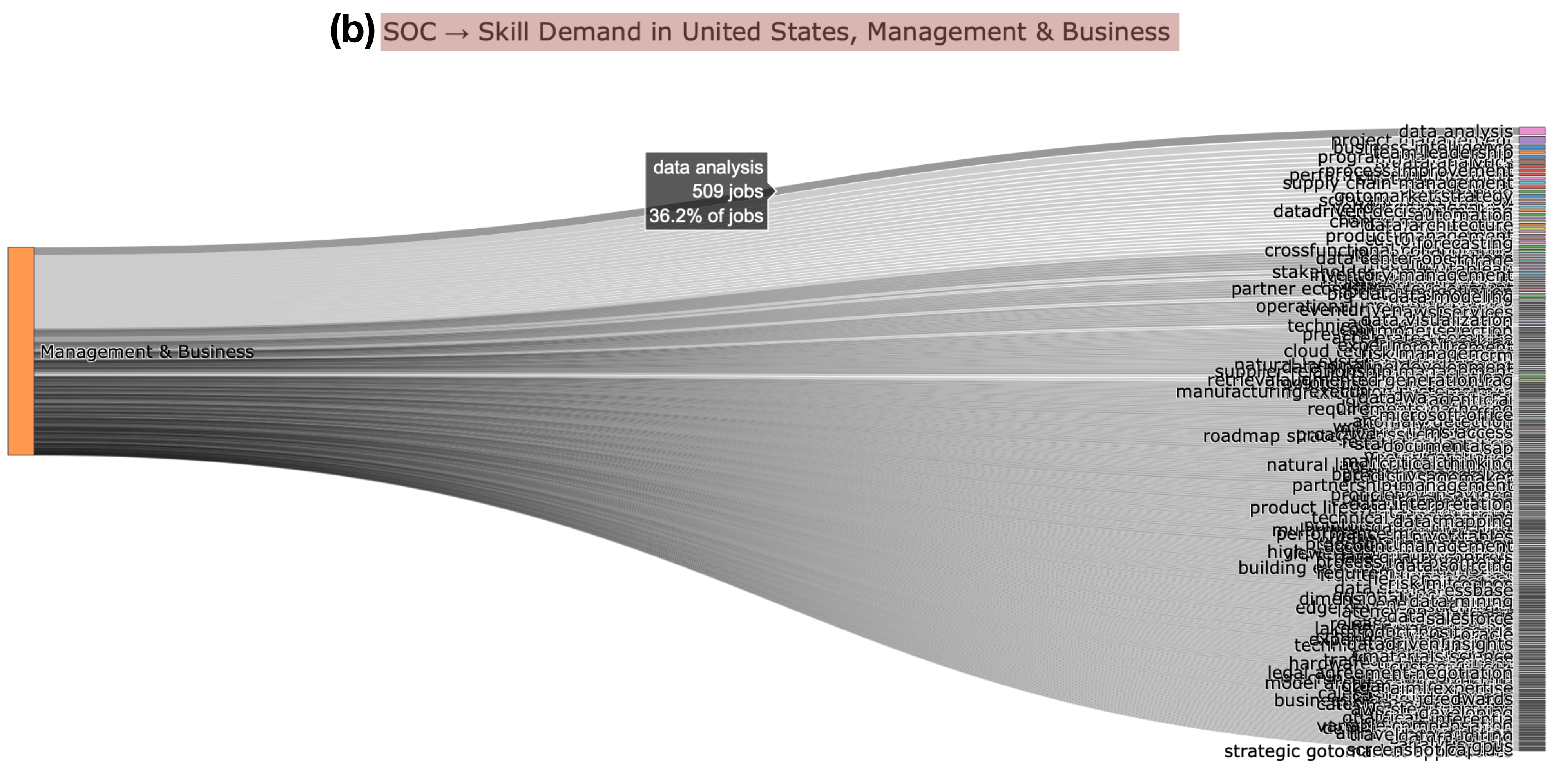}
    \end{subfigure}
    \hfill
    \begin{subfigure}[b]{0.6\textwidth}
        \centering
        \includegraphics[width=\linewidth]{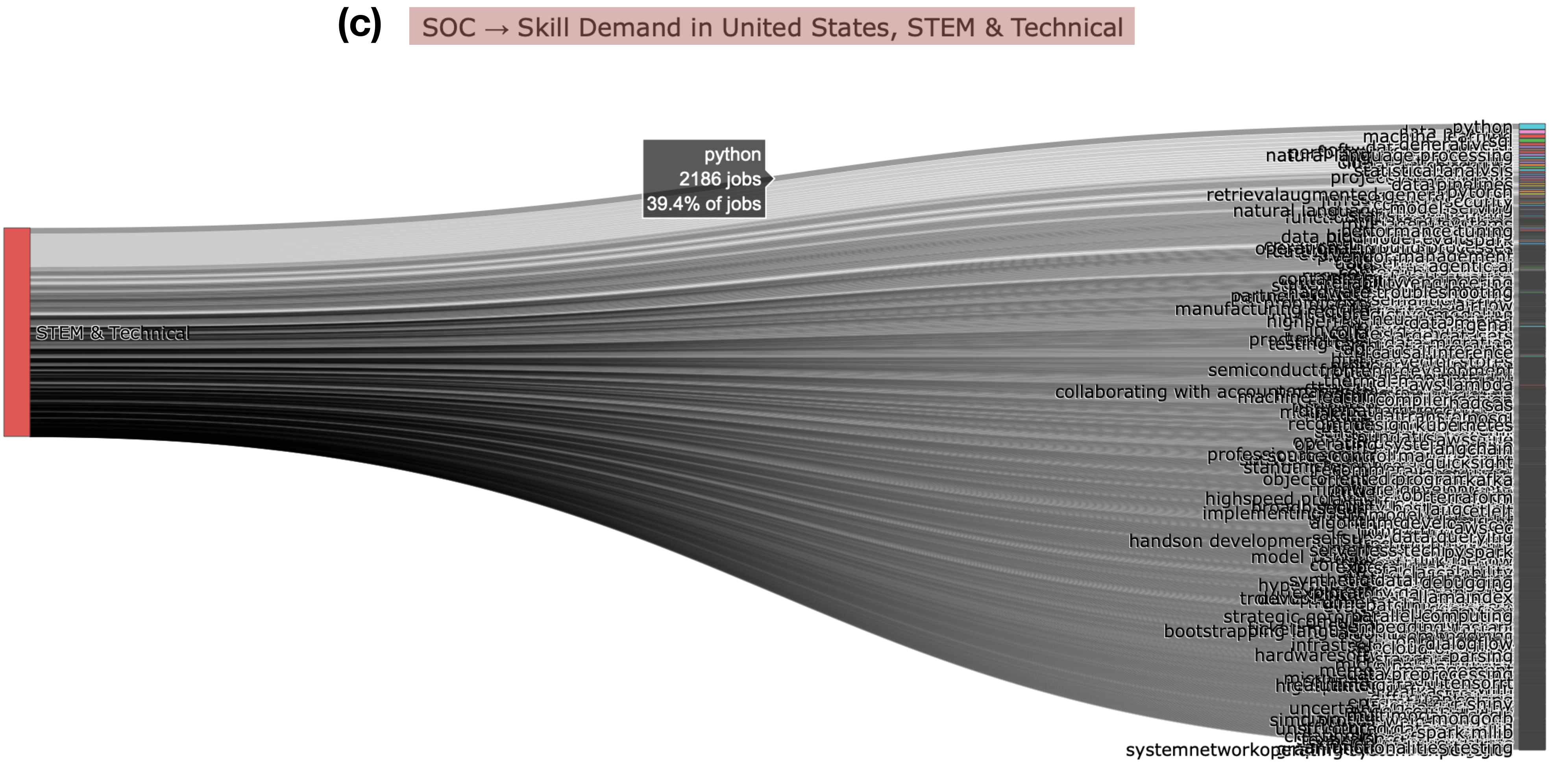}
    \end{subfigure}
        \hfill
    \begin{subfigure}[b]{0.6\textwidth}
        \centering
        \includegraphics[width=\linewidth]{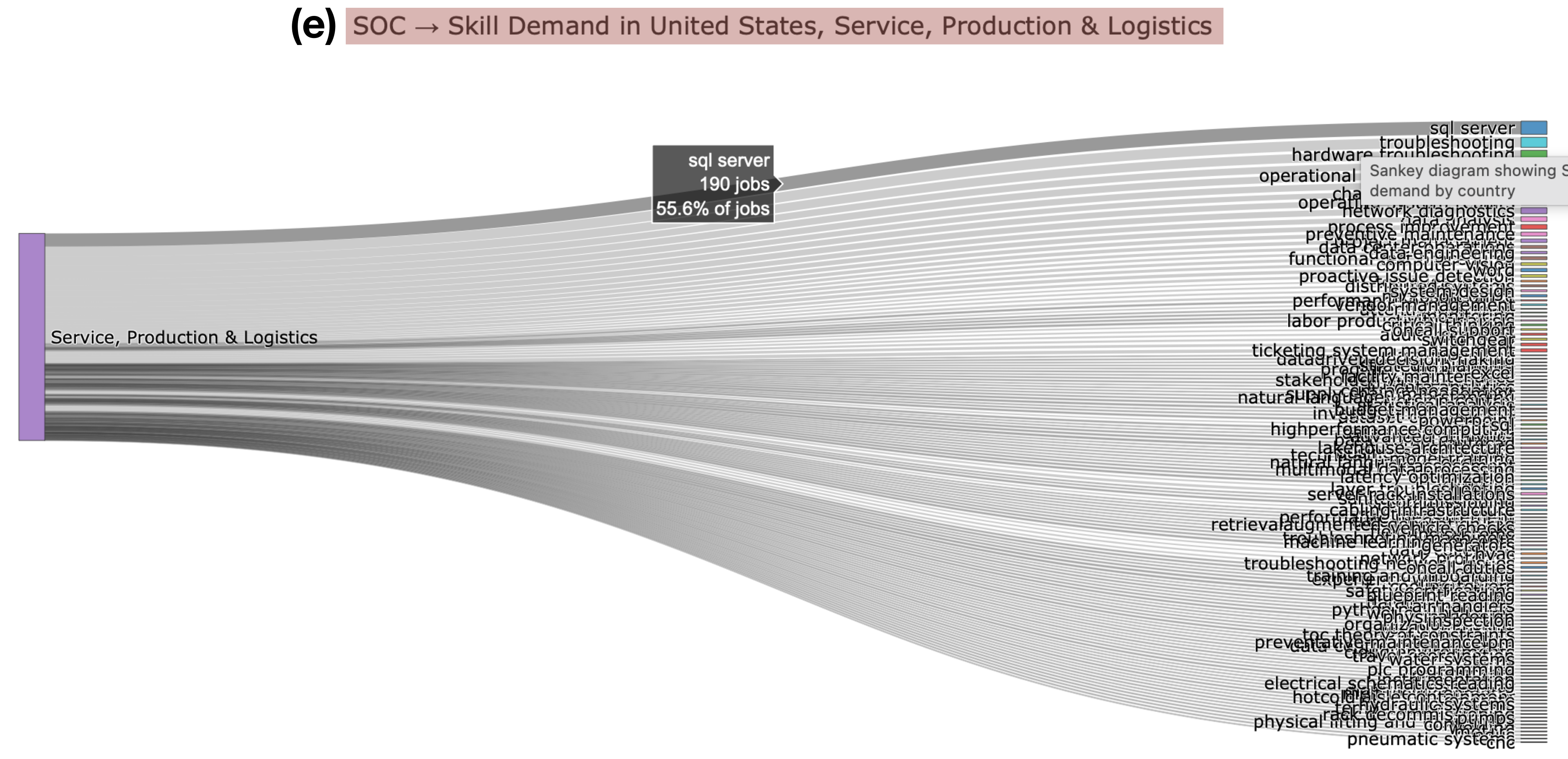}
    \end{subfigure}
       \hfill
    \begin{subfigure}[b]{0.6\textwidth}
        \centering
        \includegraphics[width=\linewidth]{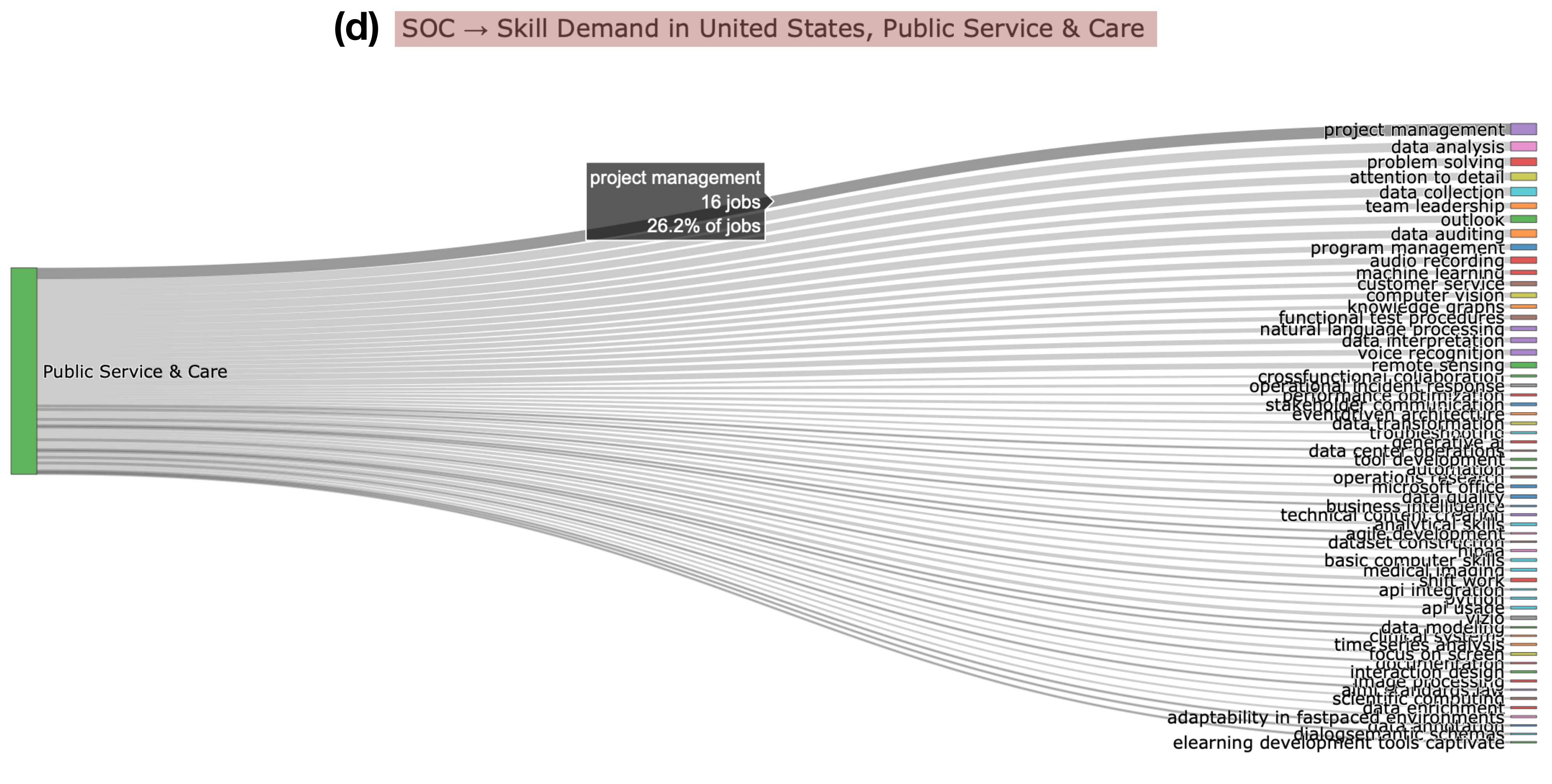}
    \end{subfigure}
    
    \caption{Distribution of skill demands by occupation sector. Sankey diagrams link each SOC occupation-cluster group to its required skills, with flow width proportional to skill frequency within the cluster.}
    \label{fig3}
\end{figure}

This aggregate demand is decomposed into its constituent competencies in Figure \ref{fig3}, presenting SOC occupation groups to skill flows in the United States as an example country. The five panels reveal pronounced heterogeneity in both the comprehensiveness and the composition of skill requirements across sectors. The STEM \& Technical cluster (Fig.\ref{fig3}c) exhibits the densest and most diversified skill structure, channeling demand into a broad portfolio anchored in machine learning, programming, and data-analytic competencies. Management \& Business (Fig.\ref{fig3}b) draws on an overlapping but narrower analytic repertoire combined with organizational and project-oriented skills. By contrast, the Health \& Education (Fig.\ref{fig3}a), Public Service \& Care (Fig.\ref{fig3}d), and Service, Production \& Logistics (Fig.\ref{fig3}e) groups display thinner and more idiosyncratic flows, with demand dispersed across a smaller and more sector-specific set of competencies. The recurrence of a common analytic necessity across the technically intensive clusters, set against markedly different skill breadth in the remaining sectors, foreshadows the dual dynamic developed through convergence around a shared data skill core coexisting with divergence across the wider occupational distribution.

\begin{figure}[!ht]
  \centering
  \includegraphics[width=1\textwidth]{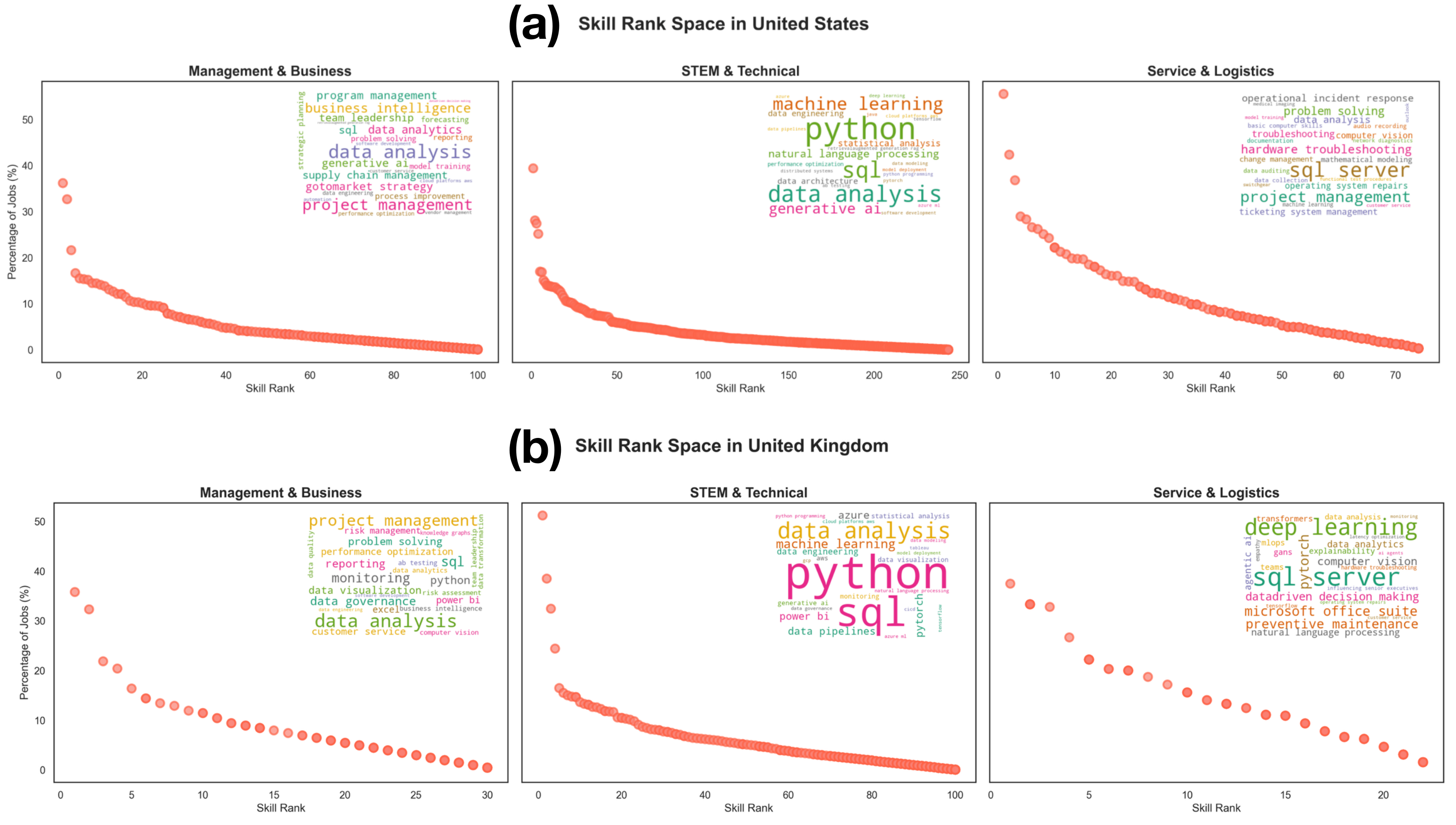}
  \caption{Skill rank space in the United States (Fig.\ref{fig4}a) and the United Kingdom (Fig.\ref{fig4}b). For each of three illustrated SOC occupation groups (Management \& Business, STEM \& Technical, Service \& Logistics), skills are ordered by descending share of relative frequency, plotting percentage of jobs against skill rank. The shape of the curve characterizes the concentration of skill demand with a steeply declining and heavy tailed curve suggesting that a few competencies account for the majority of demand, whereas a flatter curve implies more even dispersion. Overlaid word clouds flag the top ranked competencies with their size proportional to their frequency, which are dominated in the given occupational group and country.}
\label{fig4}
\end{figure}

Figure \ref{fig4} reveals the concentration of skill demand through skill rank spaces, in which competencies are ordered by their share of listings showing their rank within all skills on the horizontal axis within each cluster for the United States (Fig.\ref{fig4}a) and the United Kingdom (Fig.\ref{fig4}b). In both countries the rank distributions are strongly right skewed. It suggests that a small number of competencies account for a disproportionately large share of mentions, while the remainder form a long tail of infrequently requested skills. The head of the distribution is remarkably stable across sectors and across the two countries. Data oriented competencies, most prominently Python, SQL, machine learning, and general data analysis, dominate the upper ranks of the STEM \& Technical and Management \& Business clusters in both labor markets, while the Service \& Logistics cluster combines these with operational competencies such as SQL Server administration and hardware troubleshooting. The fact that two distinct national labor markets converge on essentially the same compact set of top ranked competencies in all sectors indicates that AI penetration is organized around an internationally transferable common skill layer rather than around nationally idiosyncratic requirements, a finding that anticipates the convergence dynamic identified in the network analysis.

\begin{figure}[!ht]
  \centering
  \includegraphics[width=1\textwidth]{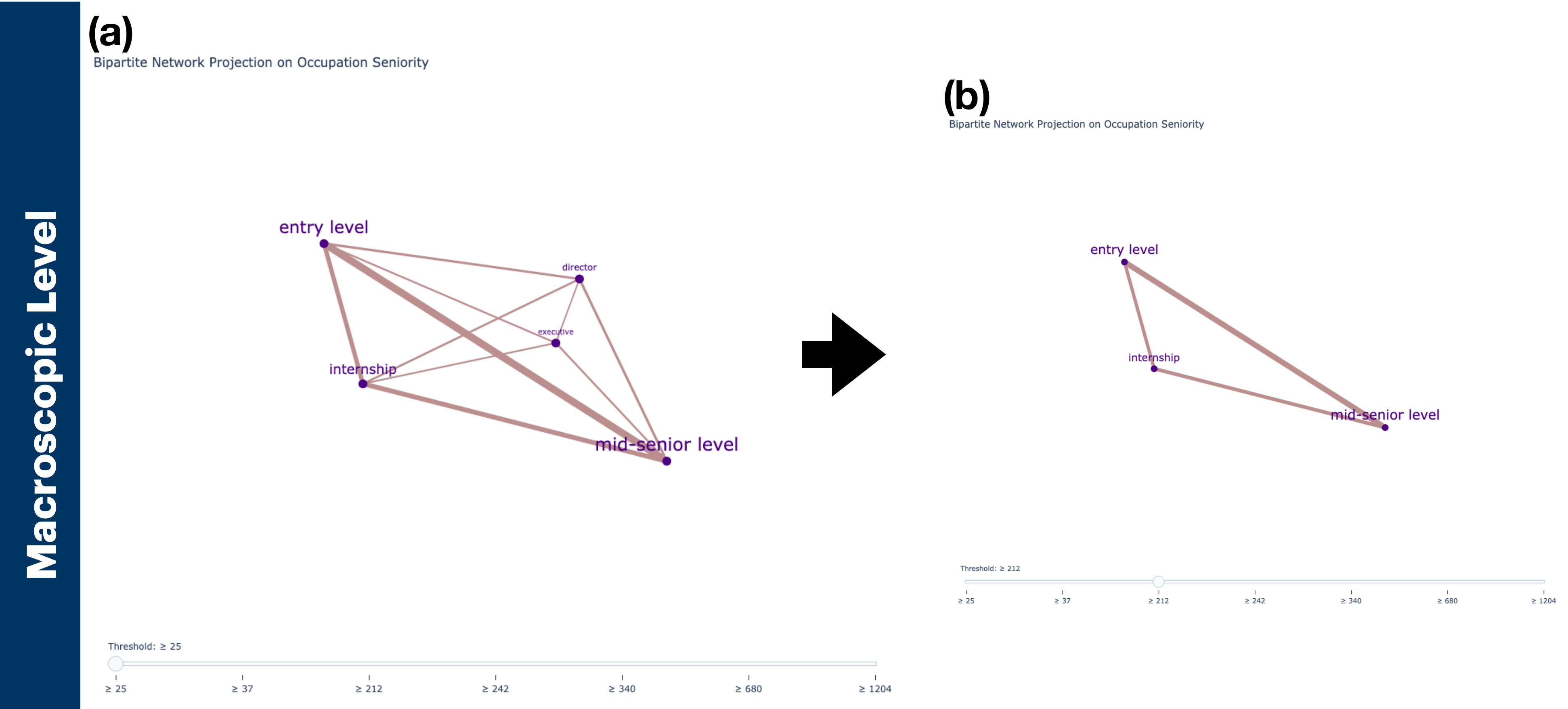}
  \caption{Bipartite network projection on occupation seniority (macroscopic level). One mode projection onto seniority levels (internship, entry, mid-senior, and director). Edges link seniority levels that share AI related skills, weighted by the number of shared skills. At a low edge-weight threshold ($\geq$ 25), all seniority levels are connected (Fig.\ref{fig5}a). At a higher threshold ($\geq$ 122), the network sparsifies (Fig.\ref{fig5}b) and the director and executive level nodes drop out, while the entry level retains the strongest ties.}
\label{fig5}
\end{figure}

The bipartite occupation–skill network projected onto the seniority dimension of occupations is illustrated in Figure \ref{fig5}, connecting seniority levels that share AI related skill requirements. In this context, the threshold is operationalized in terms of the number of shared skills between two occupations. At a low edge weight threshold ($\geq$ 25) (Fig.\ref{fig5}a), all four seniority levels from internship, entry level, mid-senior level, to director are mutually connected, reflecting a broadly shared base of AI competencies across the career ladder. As the threshold is progressively raised ($\geq$ 122) (Fig.\ref{fig5}b), the network is getting sparser and the director node is the first to detach and executive level also becomes isolated, while the entry level retains the strongest ties to the remaining structure. The persistence of entry level connectivity under increasingly stringent thresholds indicates that the most intensively and consistently demanded AI competencies are concentrated at the point of labor market entry. This pattern carries direct implications for inequality such that when employers attach substantial AI skill expectations to junior positions, AI competencies function less as a mid-career augmentation than as a precondition for entry. Therefore, it raises the effective barrier to occupational access for new entrants and for workers without prior exposure to these technologies.

\begin{figure}[!ht]
  \centering
  \includegraphics[width=1\textwidth]{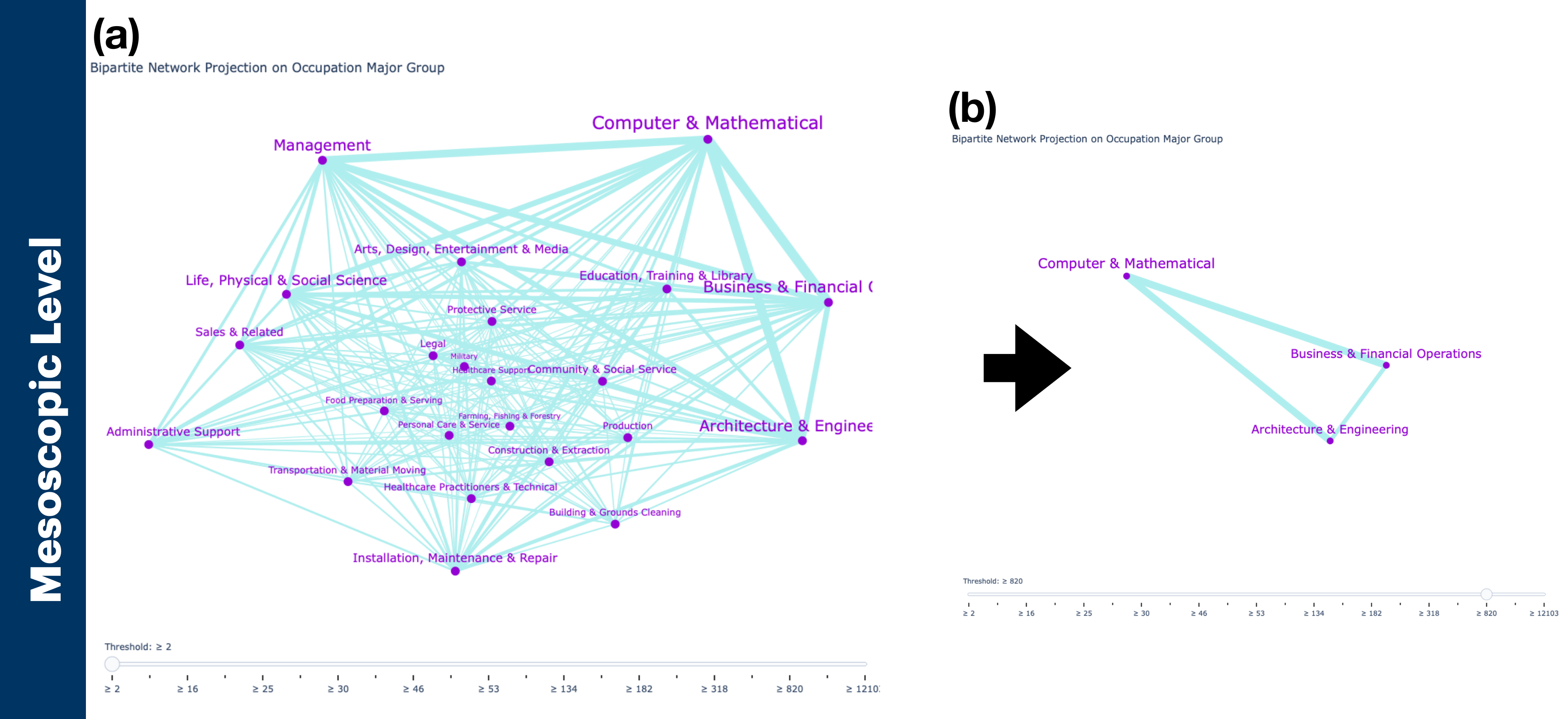}
  \caption{Bipartite network projection on SOC occupational major groups (mesoscopic level). Edges link groups that share AI related skills, weighted by the number of shared skills. At a low threshold ($\ge$ 2), the network is dense with Computer \& Mathematical occupations being central (Fig.\ref{fig6}a). Under a stringent threshold ($\ge$ 802), it reduces to a triad of Computer \& Mathematical, Business \& Financial Operations, and Architecture \& Engineering, indicating that AI convergence is channeled through the computing core and disperses toward other fields, such as business and engineering (Fig.\ref{fig6}b).}
\label{fig6}
\end{figure}

Figure \ref{fig6} projects the network onto SOC major occupation groups. The low thresholded projection ($\ge$ 2) (Fig.\ref{fig6}a) results a densely connected network, with most major groups linked by at least some shared AI competencies and Computer \& Mathematical occupations occupying a central, high strength position. As before, the threshold is defined as. the degree of skill overlap between two occupations. Applying a stringent threshold ($\ge$ 802) (Fig.\ref{fig6}b) collapses this dense web onto a compact triad comprising Computer \& Mathematical, Business \& Financial Operations, and Architecture \& Engineering occupations. The structure is consistent with a convergence dynamic that is channeled through a limited number of occupations, rather than uniformly distributed across the whole occupational system. AI skill demand is anchored in the computing core and propagates outward primarily into adjacent analytic and engineering domains, while the remaining major groups are connected only weakly and drop out under modest filtering. In substantive terms, AI driven convergence operates as a localized swing centered on Computer \& Mathematical occupations, dispersing toward business and engineering, while disappearing in other types of occupations; rather than showing a broad-based homogenization of the entire occupational structure.

\begin{figure}[!ht]
  \centering
  \includegraphics[width=1\textwidth]{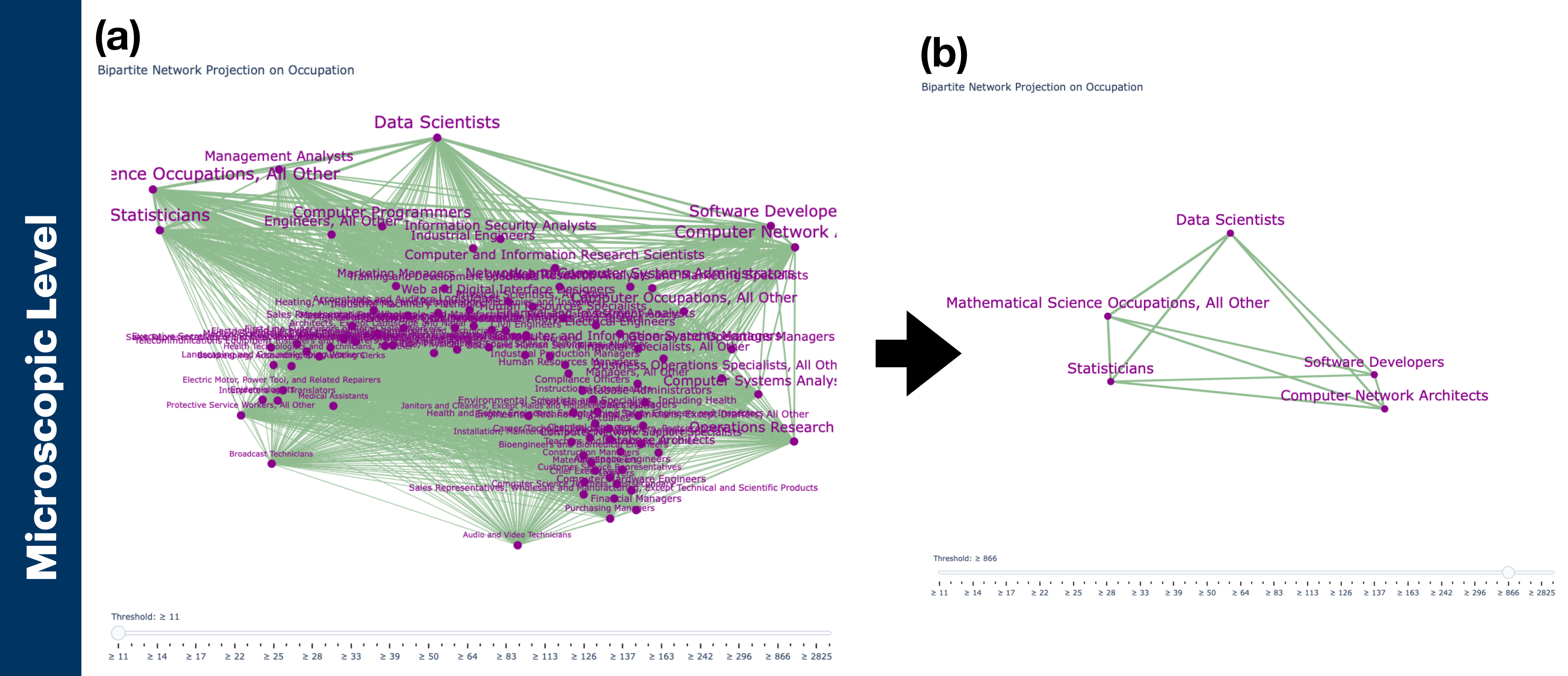}
  \caption{Bipartite network projection on individual SOC occupations (microscopic level). Edges link occupations that share AI related skills, weighted by the number of shared skills. Consistently, the threshold is expressed as the extent of skill overlap between two occupations. At a low threshold ($\ge$ 11) the network is dense, dominated by technical hubs such as Data Scientists, Software Developers, Statisticians, and Computer Network Architects (Fig.\ref{fig7}a). Under a stringent threshold ($\ge$ 866), a tight technical core remains, while managerial roles persist only through a few specialized links, indicating that technical professions require broad AI skills whereas managerial roles demand other types of capabilities (Fig.\ref{fig7}b).}
\label{fig7}
\end{figure}

Figure \ref{fig7} presents the the projection to SOC individual occupations. The full network (Fig.\ref{fig7}a) is highly interconnected and dominated by a set of high strength technical hubs, including Data Scientists, Software Developers, Statisticians, and Computer Network Architects. Under thresholding (Fig.\ref{fig7}b), a tightly knit technical core, again centered on data scientific and software occupations remains intact, while managerial and administrative occupations are retained only through a small number of specialized links. This asymmetry indicates that technical occupations converge as they are connected by a broad and overlapping portfolio of AI competencies, generating strong mutual similarity. In contrast, managerial roles draw on a narrower and more specialized subset of AI skills and thus diverge from the strongly connected hub. The coexistence of broad, shared technical demand with narrow and specialized managerial demand provides occupation level evidence for the bifurcated structure suggested by the higher level projections.

\begin{figure}[!ht]
  \centering
  \includegraphics[width=1\textwidth]{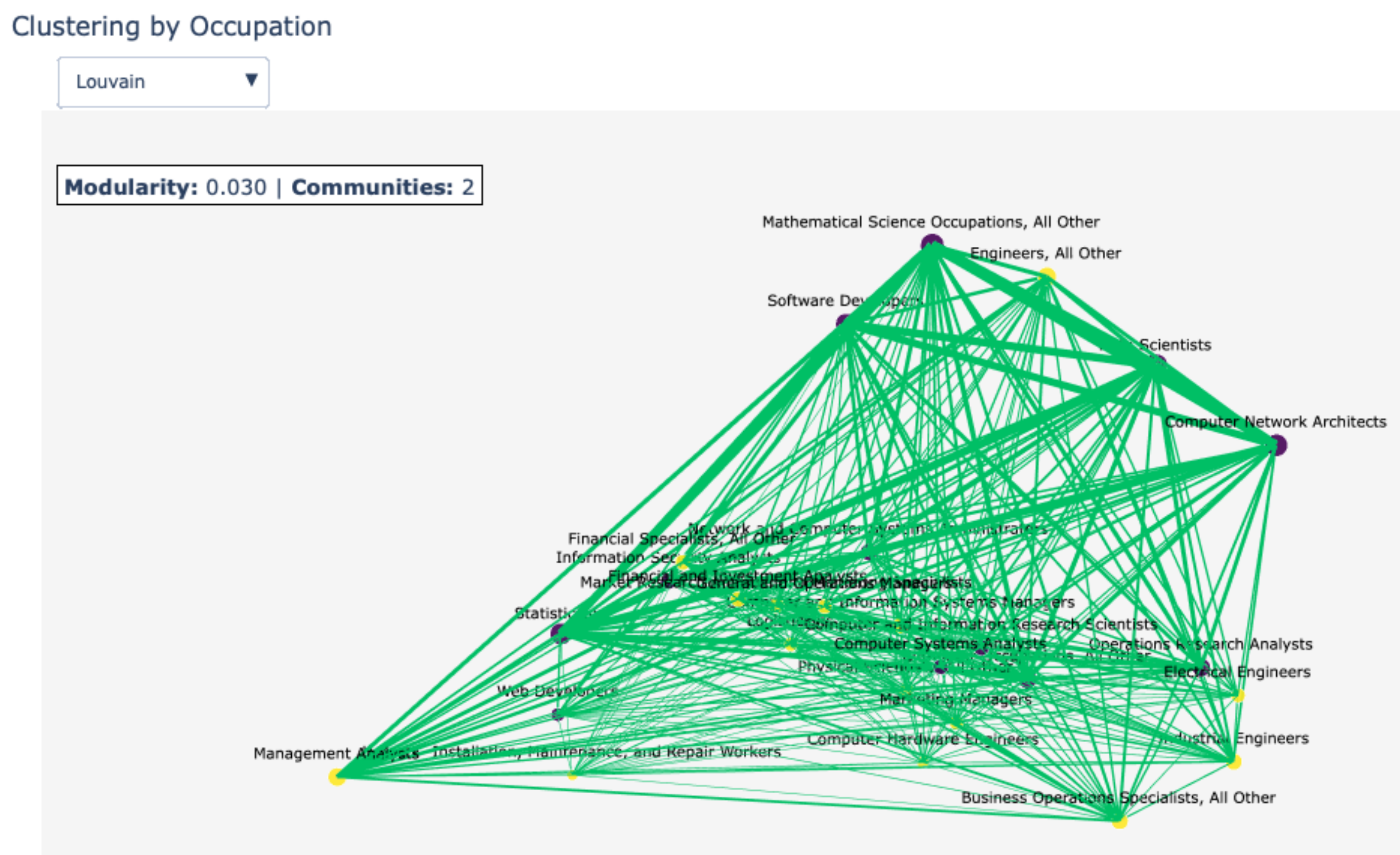}
  \caption{Community detection clustering on the projected occupation network (occupations linked by shared AI related skills) via Louvain modularity maximization. Node color denotes community membership. The algorithm detects two communities with very low modularity ($Q=0.030$), indicating weak separation between clusters and a largely shared competency pool across AI related occupations, consistent with occupational convergence.}
\label{fig8}
\end{figure}

Regarding the community structure of the occupation projection obtained with the Louvain algorithm, Figure \ref{fig8} presents the partition outcome. It yields only two communities and a very low modularity value of 0.030, indicating that the detected communities are only marginally more densely connected internally than would be expected under a random benchmark. Substantively, the near absence of strong modular separation implies that AI intensive occupations do not fragment into well separated specialized clusters but instead share a largely common pool of competencies, leaving the boundaries between communities diffuse. This weak community structure is consistent with occupational convergence within the AI exposed segment of the labor market where low level of shared competencies links otherwise distinct occupations. We note, however, that low modularity can also arise when a small number of high degree skills are required across many occupations. Interestingly, the alternative community detection algorithms namely Girvan-Newman and Leiden converged on the same two cluster solution with equally low modularity, providing additional evidence for the robustness of the identified occupational community structure as presented in the Occupation Observatory dashboard. Consequently, it reflects pervasive skill sharing.

\begin{figure}[!ht]
  \centering
  \includegraphics[width=1\textwidth]{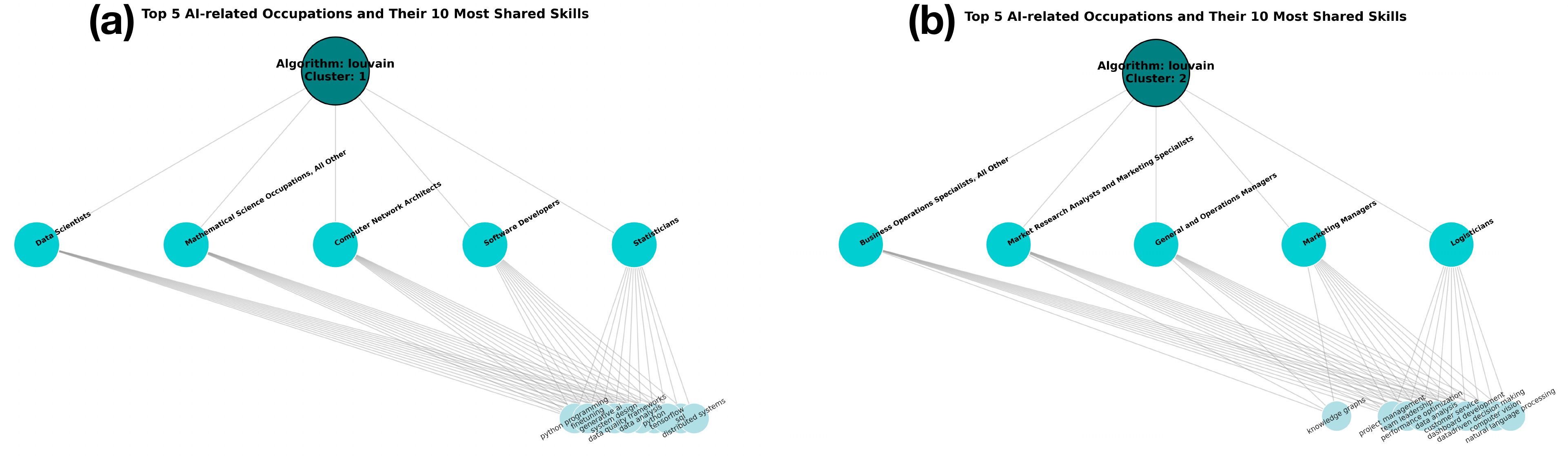}
  \caption{Top AI related occupations and their ten most shared skills by clusters derived from Louvain implementation. For each cluster, the five most AI intensive occupations (middle nodes) are shown with the ten skills most frequently shared among them (bottom nodes), as a three tier graph linkage. Fig.\ref{fig9}a and Fig.\ref{fig9}b correspond to the two communities. The large overlap of shared skills across the leading occupations of same communities indicates a common, transferable competency core, reinforcing the convergence reading from Fig.\ref{fig8}. However, the skillsets of the two communities are not overlapping, suggesting a divergence between the occupations belonging to the core group and the larger other community.}
\label{fig9}
\end{figure}

Zooming in to the configuration of each cluster, Figure \ref{fig9} displays the five most AI intensive occupations together with their ten most frequently shared skills for each of the two detected communities. The first community (Fig.\ref{fig9}a) is organized around a math, science, and model-building core. Its leading occupations are Data Scientists; Mathematical Science Occupations, All Other; Computer Network Architects; Software Developers; and Statisticians. In addition, the competencies they most frequently share are predominantly programming and model oriented python and python programming, SQL, TensorFlow, fine-tuning, generative AI, system design, distributed systems, data quality frameworks, and data analysis. The second cluster (Fig.\ref{fig9}b) comprises various occupations like business and management facing occupations such as Business Operations Specialists, All Other; Market Research Analysts and Marketing Specialists; General and Operations Managers; Marketing Managers; and Logisticians. Their shared skills blend organizational and applied analytic competencies, including project management, team leadership, performance optimization, customer service, dashboard development, data driven decision making, and knowledge graphs, together with AI application skills such as computer vision and natural language processing. In both communities the leading occupations are linked to a common set of core competencies, and the substantial overlap of these shared skills across the top occupations within the same cluster confirms that AI intensive roles are organized around a compact bundle of transferable competencies. Notably, data analysis recurs as a shared skill in both clusters, functioning as a bridging competency that spans the technical and managerial cores, whereas the remaining skills diverge sharply deep engineering and model development skills in the first community versus managerial coordination and AI enabled decision support in the second. This pattern reinforces earlier recurring pattern where generally the ties between occupations are very weak and this weakness characterize the whole network. The only exceptions are those computer, mathematical and engineering related occupations, where the shared skillset is wider and thus, links are stronger.

\section{Discussion}
According to our results, AI is not homogenizing the labor market. It is splitting it at different forces and multiple scales. The emergence of AI related skill requirements is producing neither a general convergence of occupations nor a simple intensification of specialization. Instead, it is reorganizing labor markets into a bifurcated structure. Within occupations already exposed to AI, demand converges on a compact and highly transferable core of data competencies that recurs across sectors of this group, countries, and career stages. Beyond this AI exposed core, however, convergence remains limited, concentrated within a narrow technical stratum, most strongly expressed at the point of labor market entry, and mostly reveals divergence of occupations. AI is therefore generating convergence and divergence simultaneously as convergence among technologically intensive occupations faces divergence between those occupations and the wider labor market. Furthermore, this bifurcation extends existing accounts of technological change and labor market polarization. Rather than serving as a general-purpose capability that facilitates broad occupational mobility, AI competencies appear to consolidate an already advantaged segment of the labor market while raising new barriers to entry for workers lacking prior technological exposure.

These findings both extend and qualify existing studies of AI and labor markets. The concentration of demand in high skill technical occupations is consistent with the polarization notion, in which AI complements analytical and technical labor while substituting for routine cognitive tasks \citep{hampole2025artificial, marguerit2025augmenting}. Yet our results add a structural dimension that aggregate polarization studies leave implicit since the convergence we observe operates within, rather than across, the high skill segment. The shared data skill layer is not present in other occupations to function as a general purpose bridge that would ease mobility throughout the labor market, instead, it consolidates an already advantaged technical core. This refines the optimistic reading common in policy discourse that AI competencies constitute a portable skill facilitating cross occupational transitions. Our evidence is more consistent with the alternative, in which AI skills concentrate within particular clusters of occupations and are combined with specialized knowledge to reinforce existing divisions. The persistence of these patterns across both Global North and Global South economies echoes comparative work documenting parallel polarization dynamics in different national contexts \citep{tang2022scholarly, carbonero2023impact, ganuthula2025skill} and complements regional evidence that AI exposure reshapes employment composition \citep{huang2024labor}.

The bifurcated structure we document carries direct implications for inequality, the central concern of this special issue. First, those who are opted out of the advantages of AI augmentation are more likely to remain in sectors where they are exposed primarily to the substitutional function of AI, increasing their vulnerability to displacement and reinforcing existing labor market inequalities. Second, the concentration of AI demand at the entry level reframes AI competencies less as a mid career augmentation than as a precondition for occupational access. When employers attach substantial AI skill expectations to junior positions, the effective barrier to entry rises for workers who lack prior exposure to these technologies that is in itself unequally distributed by education, institutional access, and digital infrastructure \citep{wang2024artificial}. AI penetration may therefore reproduce and harden existing inequalities at the very stage where labor market trajectories are set, a mechanism likely to be especially consequential in developing economies with constrained access to advanced training \citep{ganuthula2025skill}. Third, the limited presence of AI demand beyond technical occupations implies that the gains associated with AI related work, including the wage premia documented in prior research \citep{alekseeva2021demand}, accrue to a narrow and already advantaged occupational core, while the majority of occupations remain peripheral to AI developments. Convergence within this core thus coincides with its growing distance from the rest of the labor market, a configuration that maps onto labor market bifurcation rather than broad based upgrading. Third, the cross national consistency of the technical concentration suggests that the inequality related implications of AI penetration are structurally embedded in the organization of skill demand and are unlikely to be confined to particular national or demographic contexts \citep{huang2024labor, cranney2026global}. Reading through a sociological lens, AI here operates less as an exogenous shock than as a mechanism that channels demand along, and potentially deepens, pre-existing lines of stratification.

Reflecting on the aforementioned points, this study makes three contributions. Conceptually, it reframes the analysis of AI and labor from questions of adoption, productivity, and wages toward the structure of skill demand, operationalizing the convergence–divergence distinction through network analysis. Methodologically, it introduces an integrated, reproducible pipeline that combines collection of large scale multi country vacancy data, a two-step information extraction procedure pairing rule based parsing with zero-shot LLM classification, standardized SOC mapping, and multilevel bipartite network projection with community detection. This approach preserves the occupation–skill incidence structure while enabling comparison across levels of aggregation, from seniority and major groups to detailed occupations. Substantively, it provides cross national evidence spanning the Global North and South and delivers an interactive dashboard, \textit{The Occupation Observatory}, that makes these patterns accessible to researchers and policymakers. 

Several limitations temper these conclusions and define an agenda for further work. The analysis rests on online job postings, which capture advertised labor demand rather than realized employment, hiring, or the existing stock of workers. In addition, major platforms also potentially over represent formal, urban, and white collar vacancies. Consequently, a coverage bias likely to be more pronounced and one that may inflate the apparent concentration of AI demand in technical occupations. The detection of AI related content depends on a seed list of keywords and on zero-shot classification, which may under capture implicitly AI related tasks and introduce model specific errors that we do not formally validate against ground truth here. The data constitute a short, recent cross-section (December 2025–May 2026) and therefore describe a snapshot of the generative AI era rather than a trajectory. Our findings are descriptive as seen that the network and clustering results characterize associations in skill demand, not causal effects on employment, wages, or mobility. The low modularity we report should be interpreted as evidence of pervasive skill sharing rather than the absence of internal organization, since modularity can be depressed by a small number of high degree skills and is sensitive to algorithm and resolution choices. Finally, the analysis is demand-side. It does not observe the supply of skills or actual worker transitions, so claims about mobility remain inferential.

Future work can address these constraints along several fronts. Reducing selection bias can be obtained by expanding the seed list beyond a literature driven core through contextual embedding and co-occurrence analysis. On top of that,  the incorporation of transformer based embeddings, and topic models is promising to detect implicit AI tasks. Validating the classification step against annotated samples is also important whenever plausible. In addition, extending the data longitudinally would allow the convergence and divergence dynamics to be observed over time. Even further, linking skill demands to wage, employment, and demographic outcomes, including the gendered dimensions of AI exposure, would move the analysis from structural description toward causal assessment of how AI reshapes inequality.

\section{Conclusion}
Does the rising demand for AI related competencies is reshaping the structure of occupations toward convergence? We approached this overarching question through two objectives. The first is characterizing the relationship between required work experience and AI related skill demand across occupations, while the second is mapping the structural interdependencies between job categories and AI competencies through a multilevel bipartite network using large scale vacancy data from ten countries spanning the Global North and the Global South.

Recalling the first objective on the relationship between work experience and AI related skill demand, we find that AI competencies are demanded most intensively and most consistently at the point of labor market entry. In the seniority projection, the entry level retains the strongest connections under increasingly stringent thresholds, while the director and executive level detach. Rather than functioning as an advanced capability, AI skills increasingly operate as a precondition for occupational access, raising the effective barrier to entry for workers without prior exposure to these technologies.

Regarding the second objective on the structural interdependencies between occupations and AI competencies, the network analysis reveals a clear but uneven architecture. AI demand is overwhelmingly concentrated in STEM and technical occupations, roughly three-quarters to four-fifths of postings in every country. It rests on a compact core of data competencies led by Python, SQL, machine learning, and data analysis, that dominates the skill-rank space shown in both the United States and the United Kingdom. At the major group level of occupations, this demand is channeled through Computer \& Mathematical occupations and disperses primarily into adjacent business and engineering domains. At the occupational level, technical professions are linked by a broad, overlapping skill portfolio whereas managerial roles draw on narrower, more specialized and different capabilities. Community detection corroborates this picture by showing that the occupation network exhibits weak modular separation and resolves into a technical and model building cluster on one side and a business and management facing cluster on the other side that are bridged by a single shared competency (data analysis) but otherwise distinct in their skill requirements.

After all, these results indicate that AI penetration produces neither pure convergence nor pure divergence, but a bifurcation of labor demand. Convergence operates within the AI exposed core, where a common and portable skill layer links otherwise distinct occupations across sectors and countries. Contrastingly, divergence operates between that core and the wider labor market, into which AI demand is only present marginally. This configuration is remarkably consistent across both Global North and Global South economies, suggesting that it is structurally embedded in the organization of skill demand rather than contingent on particular national contexts.

These findings carry a clear implication for inequality. By concentrating a high value, convergent skill set within an already advantaged technical segment and attaching it to entry level access, AI penetration may reinforce existing stratification and impose new barriers to occupational mobility, particularly where the capacity to acquire technical skills is unevenly distributed in the society. Documenting these dynamics cross nationally and making them explorable through \textit{The Occupation Observatory}, this study offers an empirical foundation for monitoring how AI reshapes occupational structure and for designing the educational and labor market interventions needed to ensure that its benefits are more broadly shared.\\

\textbf{CRediT authorship contribution statement}\\
\textbf{Rafiazka Hilman}: Conceptualization, Methodology, Software, Validation, Formal analysis, Investigation, Resources, Data Curation, Writing – Original
Draft, Writing – Review \& Editing,  Visualization, Funding acquisition\\ 
\textbf{Júlia Koltai}: Conceptualization, Validation, Resources, Writing – Review \& Editing, Supervision, Project administration, Funding acquisition\\

\textbf{Declaration of competing interest}\\
The authors declare that they have no known competing financial interests or personal relationships that could have appeared to influence the work reported in this paper.\\

\textbf{Funding sources}\\
Rafiazka Hilman acknowledges support from the Momentum MSCA Programme co-funded by the European Commission through the HORIZON-MSCA-2023-COFUND programme and the Secretariat of the Hungarian Academy of Sciences (MTA) under Grant Agreement No. 101179854. Julia Koltai acknowledges funding from the
Hungarian Academy of Sciences Lendület Program: LP2022-10/2022. \\

\textbf{Declaration of generative AI and AI-assisted technologies in the manuscript preparation process}\\
During the preparation of this work the authors used GPT OSS via Ollama in order to expand the initial keyword based
search query and implement  parsing with zero-shot in the  pipeline. After using this tool/service, the authors reviewed and edited the content as needed and take full responsibility for the content of the published article.

\clearpage

\newpage

\bibliography{sample}

\end{document}


\title{Supplementary Materials}
\author{}
\maketitle

\section{Interactive Dashboard}
\label{Interactive Dashboard}
An interactive dashboard is be developed to visually illustrate the complexity of AI drift in labor market through entanglement between seniority, sector, and occupation:\\ \url{https://occupation-observatory.rafiazkahilman.com} 

\begin{figure}[!ht]
  \centering
  \includegraphics[width=1\textwidth]{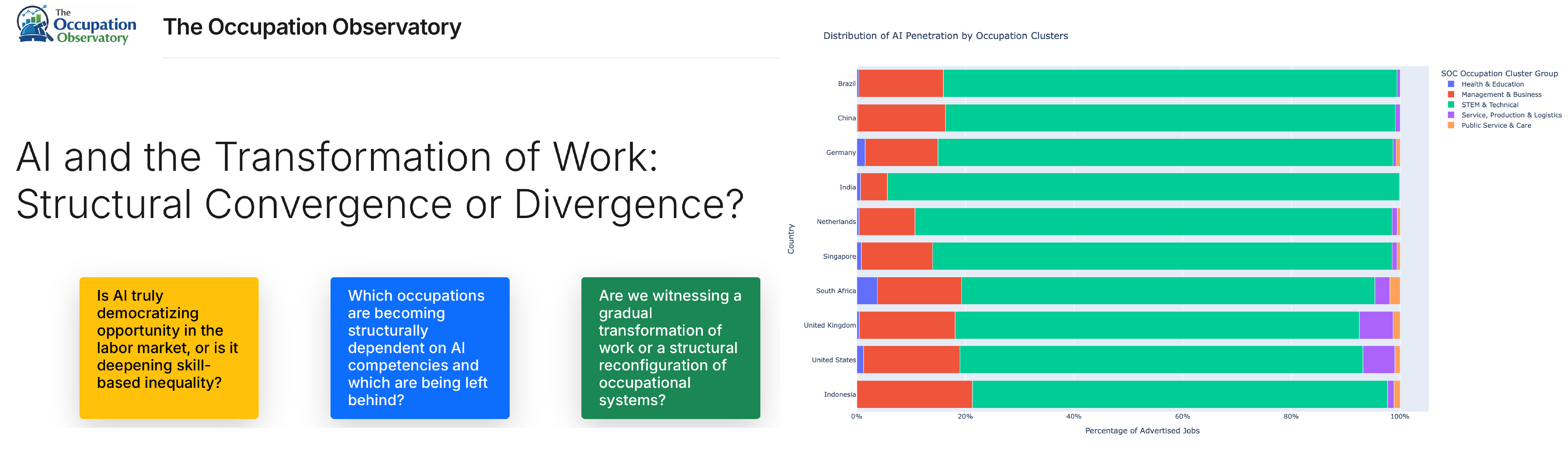}
  \caption{The dashboard interface of the Occupation Observatory.}
\end{figure}
\label{SM_A}

\section{Prompt Engineering}
\subsection{Query Expansion}
Using GPT-OSS via Ollama, we designed a prompt to expand the keyword based search query, starting from an initial set of generic terms, namely Artificial Intelligence, Data Science, Data Analytics, and Business Intelligence, as follows:
\\\

\noindent\fbox{%
    \begin{minipage}{\textwidth}
        \textit{Create a comprehensive list of AI related terms commonly used in job descriptions across major job listing platforms. The list should include technical skills, tools, frameworks, methodologies, job titles, AI concepts, infrastructure, MLOps, GenAI, LLMs, prompt engineering, AI governance, and other AI-related keywords frequently appearing in recruitment advertisements, including Artificial Intelligence, Data Science, Data Analytics, and Business Intelligence.}
    \end{minipage}%
}
\\\

This prompt enabled us to generate a comprehensive vocabulary of AI related terms commonly found in job advertisements across industries, occupations, and seniority levels, eventually improving the coverage and representativeness of the keyword based data collection process and gives these occurrences as follows:
\\\
\noindent\fbox{%
    \begin{minipage}{\textwidth}
        Machine Learning, Deep Learning, Artificial Intelligence, Python, SQL, R. Data Science, Data Analysis, Data Visualization, Automation, Predictive Analytics, TensorFlow, PyTorch, scikit-learn, Keras, Natural Language Processing (NLP), Computer Vision (CV), AWS, Azure, Cloud Services (e.g.: GCP), Docker, Kubernetes (Containerization), Big Data, Apache Spark, Hadoop, Data Engineer, ETL, Pipeline, Databases (e.g.: NoSQL), MLOps, DevOps, CI/CD, Git, GitHub, Jupyter Notebook, Agile, Scrum, API Development, Neural Networks (e.g.: CNN, RNN, Transformer), Prompt Engineering, Generative AI, Large Language Model (LLM), Model Evaluation, Data Engineering, Hadoop, Kafka, Spark, ETL, AI Ethics, Responsible AI, Governance
    \end{minipage}%
}
\\\

\subsection{Data Parsing and Classification}
Consistent with the LLM infrastructure used throughout this study, we employed GPT-OSS via Ollama to develop the following prompt for extracting structured information from job advertisements:
\\\

\noindent\fbox{%
    \begin{minipage}{\textwidth}
        \textit{\textbf{Context:} You are an expert in job analysis. Your task is to extract technical skills, soft skills, the expected start date or application deadline (if available), the corresponding Standard Occupational Classification (SOC) code, and a concise English summary from each job description. In addition, identify whether any extracted technical skills are related to Artificial Intelligence (AI), Data Science (DS), or Data Analytics (DA). }
    \end{minipage}%
}

\noindent\fbox{%
    \begin{minipage}{\textwidth}
        \textit{\textbf{Task:} From the following job description, extract a list of '$expected\_skills$' (technical skills) and '$expected\_softskills$' (soft skills). Determine whether any of the extracted technical skills are related to Artificial Intelligence (AI), Data Science (DS), or Data Analytics (DA), and return a boolean field '$has\_ai\_ds\_skill$'. Extract the '$last\_submission\_date$' in 'YYYY-MM-DD' format if explicitly stated. Otherwise, return null. Identify the most appropriate '$soc\_label$' corresponding to the Standard Occupational Classification (SOC) system (e.g.: 15-2051.00 for Data Scientists). If no suitable classification can be determined, return null. Finally, generate a '$description\_en$' field containing a concise English summary (3–5 sentences) of the job description, describing the role, key responsibilities, and required qualifications. If the original job description is not in English, first translate it into English before generating the summary.}
    \end{minipage}%
}

\noindent\fbox{%
    \begin{minipage}{\textwidth}
        \textit{\textbf{Constraint:} Provide the output in valid JSON format. The output must consist of a single JSON object with the following six keys: '$expected\_skills$' (an array of strings), '$has\_ai\_ds\_skill$' (a boolean), '$expected\_softskills$' (an array of strings), '$last\_submission\_date$' (a string in 'YYYY-MM-DD' format or null), '$soc\_label$' (a string representing the Standard Occupational Classification (SOC) code or null), and '$description\_en$' (a string or null). All entries in '$expected\_skills$' and '$expected\_softskills$' must be translated into English, regardless of the original language of the job description. The '$description\_en$' field should contain a concise English summary (3–5 sentences) describing the job role, key responsibilities, and required qualifications. If the original job description is not in English, translate it into English before generating the summary. Process the following job description: \{'$job\_description$'\}.}
    \end{minipage}%
}
\\\

This approach facilitated efficient large-scale classification of job postings without requiring task-specific examples, while ensuring that the outputs remained machine-readable and readily applicable to downstream analyses.
